\documentclass[final,twocolumn,sort&compress]{elsarticle}
\usepackage[utf8]{inputenc}
\usepackage[tbtags]{amsmath}
\usepackage{amssymb,amsthm,amsfonts,relsize,layout,indentfirst,graphicx,float,wrapfig,color,tabls}
\usepackage{boxedminipage}
\usepackage[font={bf,small},textfont={it,md,small}]{subfig}
\newcommand{\lc}[1]{\ensuremath{\mathsmaller{#1}}}
\begin{document}
%
%
	\title{Bayesian Analysis of Inertial Confinement Fusion Experiments at the National Ignition Facility}
	\tnotetext[t1]{This work performed under the auspices of the U.S. Department of Energy by Lawrence Livermore National Laboratory under Contract DE-AC52-07NA27344.}
	\author[llnl]{J.A.~Gaffney\corref{cor1}}
	\ead{gaffney3@llnl.gov}
	\author[llnl]{D.~Clark}				
	\author[llnl]{V.~Sonnad}
	\author[llnl]{S.B.~Libby}
	\cortext[cor1]{Corresponding author}
	\address[llnl]{Lawrence Livermore National Laboratory, 7000 East Ave, Livermore, CA 94550}
%
%
%
%
	\begin{abstract}
	We develop a Bayesian inference method that allows the efficient determination of several interesting parameters from complicated high-energy-density experiments performed on the National Ignition Facility (NIF). The model is based on an exploration of phase space using the hydrodynamic code HYDRA. A linear model is used to describe the effect of nuisance parameters on the analysis, allowing an analytic likelihood to be derived that can be determined from a small number of HYDRA runs and then used in existing advanced statistical analysis methods. This approach is applied to a recent experiment in order to determine the carbon opacity and X-ray drive; it is found that the inclusion of prior expert knowledge and fluctuations in capsule dimensions and chemical composition significantly improve the agreement between experiment and theoretical opacity calculations. A parameterisation of HYDRA results is used to test the application of both Markov chain Monte Carlo (MCMC) and genetic algorithm (GA) techniques to explore the posterior. These approaches have distinct advantages and we show that both can allow the efficient analysis of high energy density experiments.
	\end{abstract}
%
%
	\maketitle
%
%
\section{Introduction}
High energy density experiments conducted on large laser facilities are often highly complex in every way. The physics of the target's interaction with the laser, its evolution with time, the relevant material properties at the conditions of interest, and the relation between target properties and observable quantites all require significant effort to describe, and all involve approximations. This is particularly true of so-called `integrated' experiments, such as the inertial confinement fusion (ICF) effort underway at Lawrence Livermore National Laboratory \cite{moses10,glenzer12}, in which extremely complex targets probe all aspects of the ICF approach to nuclear fusion.\par
In general the only way that such complex systems can be accurately understood is through computer simulation. These simulations take a range of inputs desribing the target, laser system, and input physics and provide a description of the target evolution. This can then be mapped onto the output of experimental diagnostics. The approach represents a nonlinear, `black box' transformation from the uncertain input physics to the measured quantites; the inversion of this to provide information about difficult physics can be a very difficult statistical problem. In this paper we discuss the application of Bayesian statistics \cite{sivia} to the solution of this problem, and in particular to the inference of interesting physical parameters from an ICF experiment performed at the National Ignition Facility (NIF).\par
The Bayesian approach is a well developed statistical method that can be formulated to give powerful results for the inference of interesting quantities \cite{hanson04,dodt08,broomhall10}, or alternately to provide a rigorous framework for the design of new experimental methods \cite{sacks89,chaloner95,dreier08}. Bayesian theory includes the prior knowledge of the system explicitly and so allows the evolution of interesting quantities to be charted as data accumulates from various sources. The numerical application of the theory is usually through Markov chain monte carlo techniques \cite{green95}; in the below we develop a method of incorporating the advantages of the Bayesian theory into genetic algorithm approaches which can be parrallelised very easily. These numerical methods have been applied in almost all areas of science for data analysis, including fusion studies, high energy density physics experiments \cite{golovkin00}, and hydrodynamic simulations \cite{hanson04,williams06}. The example chosen here includes aspects of all of these types of experiment.\par
In the following section the Bayesian approach is set out. Particular attention is paid to the interpretation of Bayes' theorem as an information processing rule, since this can be very useful in understanding the evolution of physical models. The relationship between Bayesian methods and the common $\chi^{2}$ methodology is discussed, which allows a generalised $\chi^{2}$ to be defined that takes into account the effect of nuisance parameters in an experiment, and of prior knowledge of the system. This is applied to a NIF `convergent ablator' experiment, and used to show the variation in the expected carbon opacity as these Bayesian factors are introduced. Finally, the MCMC and GA approaches are applied to the same analysis and discussed.\par
%
%
\section{The Bayesian approach}
In the Bayesian approach real numbers are used to describe the level of belief that a statement is true. A set of intuitive axioms govern these numbers, which coincide with the axioms of the usual probability theory. As such, Bayesian statistics can be developed as a re-intepretation of frequentist statistics; rather than the usual understanding that the probability $P(X=x)={\rm Lim}_{N\rightarrow \infty} \frac{N(X=x)}{N}$, $P(X=x)$ becomes the belief that the statement $X=x$ is true. This belief depends on some background information $\mathbb{I}$, a dependance that is expressed using the conditional probability $P(X=x \vert \mathbb{I})$.\par
An experiment has the effect of updating the background information with some observation $D=d$, so that after the experiment the belief becomes $P(X=x \vert D=d,\mathbb{I})$. The analysis of this updated probability forms the basis of Bayesian analysis. The axioms of the theory allow the important Bayes' theorem to be derived,
	\begin{equation}
		P(x \vert d,\mathbb{I}) \propto P(d \vert x, \mathbb{I}) P(x \vert \mathbb{I})
		\label{eq:bayes_theorem}
		{\rm ,}
	\end{equation}
which relates the probability after the observation of $D$ (the \emph{posterior}) to the probability before (the  \emph{prior}) through the probability of observing the data when the value of $X$ is assumed, $P(d \vert x, \mathbb{I})$ (the \emph{likelihood}). The terms in equation \eqref{eq:bayes_theorem} can be viewed as functions of any of their arguments; the study of $P(x \vert d,\mathbb{I})$ as a function of $x$ allows the value of $x$ to be inferred from data, whereas treating $P(x \vert d,\mathbb{I})$ as a function of $d$ allows the significance of particular observations to be explored. The latter of these allows the design of experiments by ensuring that an experiment collects the most significant data possible.\par 
Bayes theorem can be viewed as rule for describing the flow of information in an experiment \cite{zellner88}. The \emph{information} associated with an observation of the random variable $Z$ is
	\begin{equation}
		I(Z=z) = -{\rm ln}P(Z=z)
		\label{eq:information}
		~{\rm ,}
	\end{equation}
where the above is measured in `nats'. The more usual unit of `bits' is found using a base 2 logarithm. The information associated with an unlikely observation is greater than that with an unsurprising result. Bayes' theorem becomes
	\begin{equation}
		I(x \vert d,\mathbb{I}) = I( d \vert x,\mathbb{I}) + I (x \vert \mathbb{I}) + {\rm constant}
		\label{eq:bayes_theorem_information}
		~{\rm ,}
	\end{equation}
and so so the effect of an experiment is to add the new and prior informations. The information defined in equation \eqref{eq:information} is evaluated for a given observation result $z$; when this is not known (as in problems of experimental design) it is necessary to consider the expectation value of the information, the \emph{information entropy}
	\begin{equation}
		H[P] = -\int P(z){\rm ln} P(z) dz
		\label{eq:information_entropy}
		~{\rm ,}
	\end{equation}
which is now a property of the distribution $P$ only. The \emph{cross entropy} between two distributions $P$ and $Q$,
	\begin{equation}
		H\left[ P \Vert Q \right] = -\int P(x){\rm ln} Q(x) dx
		\label{eq:information_cross_entropy}
		~{\rm ,}
	\end{equation}
measures the common information in the two distributions. This is particularly useful since it is conserved in Bayes' theorem \cite{zellner88}. Using the above definition equation \eqref{eq:bayes_theorem} can be written as
	\begin{equation}
		H\left[ P(x\vert d,\mathbb{I}) \left\Vert \frac{ P(x\vert d,\mathbb{I}) }{ P(x\vert \mathbb{I}) } \right.\right] = H\left[ P(x\vert d,\mathbb{I}) \Vert P(d\vert x,\mathbb{I}) \right]
		\notag
		~{\rm ,}
	\end{equation}
or
	\begin{equation}
		H_\lc{KL}\left[ P(x \vert d,\mathbb{I}) \Vert P(x \vert \mathbb{I})\right] = H\left[ P(x\vert d,\mathbb{I}) \Vert P(d\vert x,\mathbb{I})\right]
		\label{eq:KL_distance}
		~{\rm ,}
	\end{equation}
so that the increase in cross entropy between the posterior and prior (the \emph{Kullback-Leibler distance}) is equal to the cross entropy between posterior and likelihood. Experimental design problems often maximise $H_{KL}$ to maximise the effectiveness of an experiment. Using the above definitions it has been shown that Bayes' theorem is the only information processing rule that exactly conserves information cross entropy \cite{zellner88}.\par
The above definitions are all useful forms of the same rule; the information that a given experiment provides is described by the likelihood function $P(d \vert x, \mathbb{I})$. Data analysis is then the analysis of the likelihood function for a given experiment and model, and standard models can be understood as special cases of the likelihood and prior. For example, it is common to use a normally distributed likelihood
	\begin{equation*}
		P(d \vert x,\mathbb{I}) \propto e^{-\frac{1}{2}\left( \frac{d - \hat{d}(x)}{\sigma}\right)^{2}}
		~{\rm ,}
	\end{equation*}
where $\hat{d}(x)$ is a model for the data given the parameter $x$; in this case the information associated with an observation is the usual $\chi^{2}$ function \cite{sivia}. The $\chi^{2}$ minimisation method is then equivalent to finding the value of the parameter $x$ that minimises the information provided by the experiment (that is, the parameter for which the observation is the least surprising). For a constant prior probability $P(x\vert \mathbb{I})$ (often used to describe complete ignorance) this is in turn equivalent to finding the value of $x$ that has the largest posterior probability. Hence the $\chi^{2}$ minimisation method can be seen to contain two approximations; the neglect of the structure of the posterior, and the neglect of prior information.\par
The prior information plays an important role in Bayesian analysis. In general it contains all information that is available before the experiment; expert opinion, the results from previous experiments, etc. Its inclusion allows otherwise very difficult problems to be tackled, and allows the belief in a model or parameter to evolve over time by using the posterior from one experiment as the prior of another. As such data analysis can include the results from entire campaigns of experiments common in modern science, and these experiments need not have the same design or be performed by the same people. All that is required is an expression for the likelihood function for a given experimental design.\par
%
%
\section{Calculation of the likelihood function from large scale hydrodynamic simulations}
In this work we aim to investigate the Bayesian analysis of inertial confinement fusion experiments performed at the National Ignition Facility. The usual analysis procedure is to use radiation-hydrodynamic simulations to provide post-shot analysis; measured quantities like laser power are used as inputs for simulations, the outputs of which are compared to other measured quantities. This comparison is used to gain information about the physics used by simulations, and other unmeasured properties of the experiment.\par
We consider NIF shot N110625, a `convergent ablator' experiment designed to mimick the implosion dynamics of a full ignition shot. This shot is of interest since large-scale post shot simulations have been performed in the past, in which a large number of target parameters were varied. This dataset will serve to benchmark the smaller set of simulations used in this study.\par
\subsection{HYDRA Simulations}
The eventual aim is that advanced numerical methods such as Markov-Chain Monte Carlo (MCMC) routines, or Genetic Algorithms (GA), are used to perform efficient analysis of NIF data. To acheive this a robust platform for running hydrodynamic simulations is required, and the feasibility of such large-scale simulations must be investigated. To that end a set of fortran wrapper routines have been developed that allow the radiation-hydrodynamic code HYDRA to run as a subroutine to a generic MCMC or GA code. These routines take a set of values of various input parameters (opacity, equation of state and drive modifiers, etc.) and return various measurable characteristics of the implosion. Modification of input physics databases, and extraction of implosion dynamics, is performed using existing scripts to ensure that returned values are consistent with those commonly used in NIF data analysis. Calculations are done in parallel, allowing an analysis code to consider several tens of thousands of HYDRA simulations in one job allocation.\par
\begin{figure}
		\centering
		\includegraphics[scale=1.0]{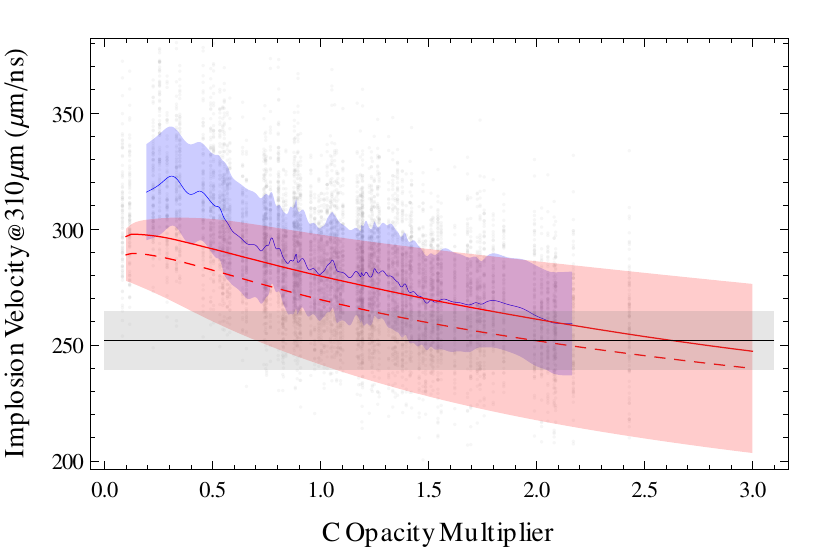}
		\caption{The implosion velocity as a function of carbon opacity multiplier for HYDRA simulations of N110625. Blue and red regions show one standard deviation about the mean of previous design simulations and this work respectively. Black shading indicates the experimental error bar.}
		\label{fig:me_vs_dc}
	\end{figure}

In this work a grid of 1024 simulations of a convergent ablator implosion has been generated, in which a constant multiplier is applied to the absorption opacity of carbon and to the X-ray drive. A representation of the results is shown in figure \ref{fig:me_vs_dc}, in which the velocity of the centre of mass of the fuel is plotted as a function of carbon opacity multiplier (averaging over all other parameters). Black points show the results of the earlier parameter scan, with the blue line and shading representing a moving average of the mean and standard deviation of those results. The mean and standard deviation of the results of this work are shown in red. The solid line represents the nominal (unmodified drive) case, and the broken line shows the drive-averaged case. The large difference between these lines demonstrates the difficulty in increasing implosion velocity in these targets. Finally, the black region shows the experimental value \cite{hicks_pc1}. \par
The data in figure \ref{fig:me_vs_dc} agree well for the nominal case. The large number of parameters in the original data results in statistical noise in these results, the non-overlapping set of modifications made to simulations (original simulations focused more on changes to the capsule design than input physics), and the differences between the sampling of parameter space mean that the averages do not agree. The good agreement between the current work's nominal case and the previous results demonstrates that the HYDRA simulations used here are reasonable. It should be noted, however, that the simple modifications to the drive used here do not reflect the current understanding of the X-ray drive in ICF experiments, rather they were chosen as the drive multiplier is likely to correlate with the opacity multiplier.\par
\subsection{Calculating the Likelyhood}
We will use three measured results in the calculation of the likelihood; the implosion velocity when the centre of mass of the fuel reaches a radius of 310$\mu$m, the time at which this radius is reached, and the fraction of ablator mass still remaining at this time \cite{hicks_pc1}. Experimental data for these are available, and they can be easily extracted from HYDRA simulations. There are no simulated points that are able to match all three experimental values, making a statistical approach essential. \par
\begin{figure}
		\centering
		\includegraphics[scale=1]{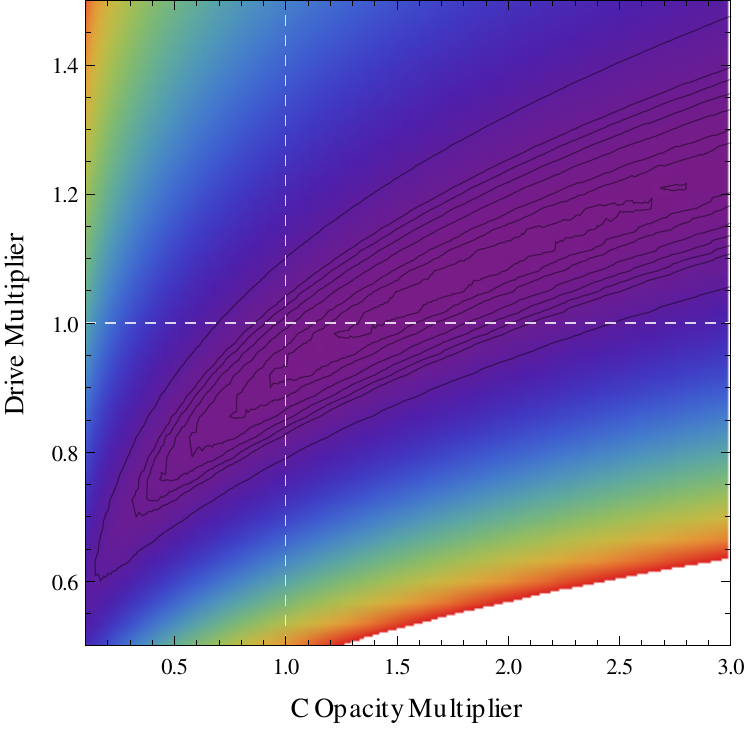}
		\caption{The likelihood as a function of simulation parameters, calculated using a multivariate normal distibution. Contour lines are logarithmic and centered on the minimum information parameters}
		\label{fig:normal_likelihood_info}
	\end{figure}

 
As discussed in the previous section, a simple model for the likelihood is an uncorrelated multivariate normal distribution. The information associated with the likelihood in this case, as a function of the two parameters we wish to infer, is plotted in figure \ref{fig:normal_likelihood_info}. The most likely values of the drive and opacity multipliers occur at the minimum point of this surface, when the data are the least surprising.
If each of the 3 experimental points are considered individually, each one defines a range of values of opacity and drive multipliers that best match the experimental result. The minimum information point in figure \ref{fig:normal_likelihood_info} results from the superposition of the three data points, giving rise to a broad range of (approximately) equally likely values of the simulation input parameters.\par 
The assumption of a normal distribution for the calculation of the likelihood is based on assumptions about the nature of the \emph{experimental} error. Simulations are treated as deterministic. In practice this is not the case; the values of input parameters will have a probability distribution as well (reflecting the accuracy with which they are calculated, for example) and the resulting distribution of code outputs will change the shape of the likelihood function. Proper inclusion of all sources of error rapidly increase the parameter space for simulations and so it is advantageous to use an approximate method for some parameters. For this we use a normal linear model \cite{chaloner95}, which we develop below.\par
Consider the likelihood function for a simulation with two inputs as before, but where one is a `nuisance' parameter. If the nuisance parameter is normally distributed, and the response of the simulation to variations in the parameter is linear, then the likelihood function is a normal distribution with a modified width (see below). This linear response method can be used to include the effect of a large number of nuisance parameters with a greatly reduced number of simulations. The potential accuracy of this approach is improved by the fact that input physics models, target fabrication, etc. are highly developed for ICF applications making the expected variations in nuisance parameter values small.\par
To develop the linear response model, consider a normal approximation in which the likelihood is given by the multivariate distribution with covariance matrix $\Lambda_{d}$,
	\begin{equation*}
		P(d \vert x,\mathbb{I}) =  \frac{e^{-(d-\hat{d}(x))^{T} \Lambda_{d}^{-1} (d-\hat{d}(x))}} {\sqrt{|\Lambda_{d}|(2\pi)^{n_{d}}}}
		~{\rm .}
	\end{equation*}
Splitting the simulation input vector $x$ into the set of interesting parameters $x^{\prime}$ and nuisance parameters $y$, introducing a second multivariate normal distribution (of covariance $\Lambda_{y}$) to describe the variations in nuisance parameters, and treating the response of a simulation as linear,
	\begin{equation}
		\hat{d}(x)=\hat{d}(x^{\prime},y) \simeq \hat{d}(x,\bar{y}) + A(y-\bar{y})
		~{\rm ,}
		\label{eq:linear_response}
	\end{equation}
the likelihood function can be found analytically. Using $P(d | x^{\prime}) = \int P(d | x^{\prime},y) P(y) dy$, the result is
	\begin{equation}
		P(d \vert x^{\prime},\mathbb{I}) = \frac{e^{-(d-\bar{d}(x^{\prime}))^{T} \left[ \Lambda_{d}^{-1}-\beta^{T}\beta\right] (d-\bar{d}(x^{\prime}))}}{\sqrt{(2\pi)^{n_{d}}|\Lambda_{d}||\Lambda_{y}||\alpha^{T}\alpha|}} 
		~{\rm ,}
		\label{eq:likelihood_marginal}
	\end{equation}
where $\bar{d}(x^{\prime})$ is the nominal simulation result $\hat{d}(x^{\prime},\bar{y})$ and the matrices $\alpha$ and $\beta$ satisfy the equations
	\begin{align}
		\alpha^{T}\alpha &=A^{T}\Lambda_{d}^{-1}A+\Lambda_{y}^{-1} \notag \\
		\beta^{T}\alpha &= \Lambda_{d}^{-1}A \notag
		~{\rm .}
	\end{align}
In this approximation the likelihood is a normal distribution with a modified covariance matrix. The modification is not diagonal, and so the effect of the simulation response is to distort and rotate the likelihood; the rotation is towards an axis determined by the strength of the simulation response to the various nuisance parameters and by the expected variation in those parameters. In the case of current interest, in which experimental data have already been observed, the information associated with the observation $d$ (equation \eqref{eq:bayes_theorem_information}) is modified to become
	\begin{equation}
		\begin{split}
		I(d|x^{\prime},\mathbb{I})=& \sum_{i} \frac{(d_{i}-\bar{d}(x^{\prime})_{i})^{2}}{\sigma_{i}^{2}} - {\rm ln}P(x | \mathbb{I}) \\
			 & - (d-\bar{d}(x^{\prime}))^{T} \beta^{T}\beta (d-\bar{d}(x^{\prime})) \\
			 & +\frac{1}{2}{\rm ln}\left(|\Lambda_{y}||\alpha^{T}\alpha|\right)
		\end{split}
		~{\rm ,}
		\label{eq:information_likelihood_marginal}
	\end{equation}
which defines a modified $\chi^{2}$ function that takes into account prior information and the effect of nuisance parameters. These considerations have an important effect on the information associated with a given measurement, and therefore the results of any analysis.\par
%
%
\section{Application to a NIF Convergent Ablator implosion}
The effect of nuisance parameters on N110625 can be included in the above way. We use previous design calculations to populate the response matrix $A$, treating variations in capsule dimensions and chemical composition as nuisance parameters. These are allowed to vary according to their specified manufacturing tolerances \cite{haan11}. The modification to the likelihood by these parameters is quite significant; the power of the linear response model is that once the modifications (in the form of the new covariance matrix) are known, subsequent analyses do not need to explicitly include the effect of nuisance parameters at all.\par
\begin{figure}
		\centering
		\subfloat[Nuisance Parameters Ignored]{
			\includegraphics[scale=1.0]{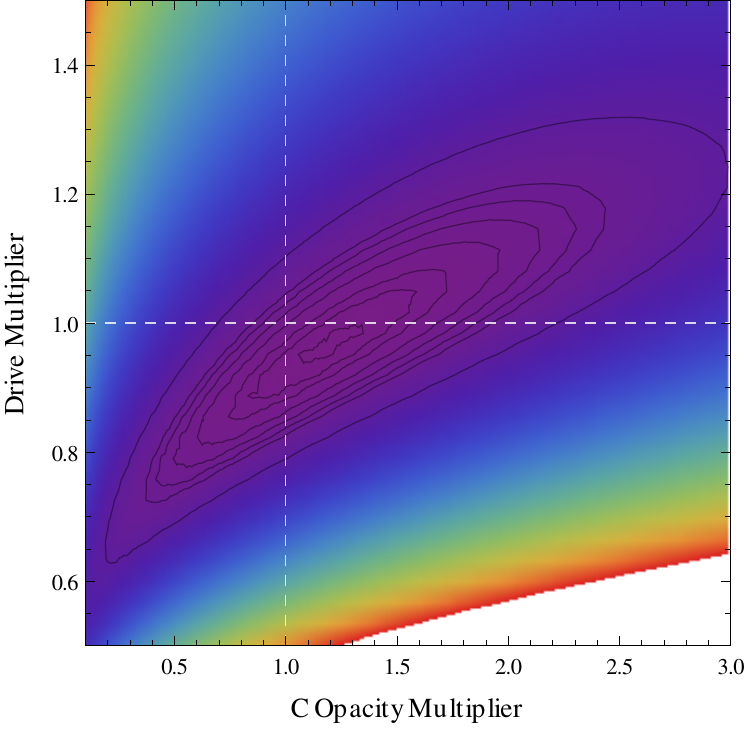}
			\label{subfig:no_marginal}
		}
		\vspace{1cm}
		\subfloat[Nuisance Parameters Included]{
			\includegraphics[scale=1.0]{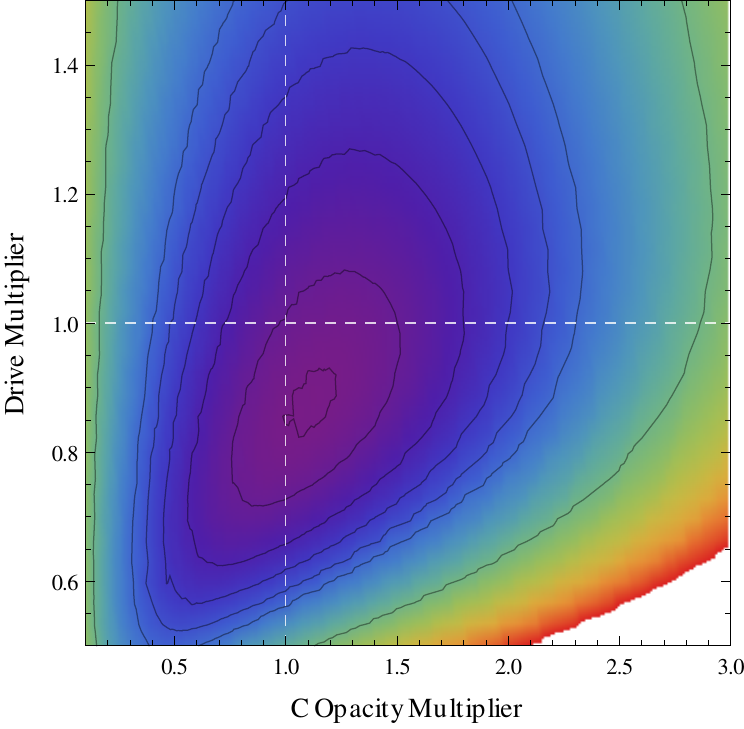}
			\label{subfig:marginal}
		}
		\caption{The information contained in the posterior for an analysis of NIF shot N110625. Figure (a) shows the result when variations in the capsule dimensions and atomic composition are neglected, and (b) shows the result when these are included in a linear approximation. Both parameters have a normal prior of standard deviation 0.1. Contour lines show constant fractions of the peak information. Nuisance parameters have the effect of producing a more localised peak in the posterior, however this peak is broader.}
		\label{fig:final_info}
	\end{figure}

Given knowledge of the prior, from the work in the previous sections and the results of HYDRA simulations, we can infer the drive and carbon opacity multipliers from the described data, including prior information and the effect of nuisance parameters. The results are shown in figure \ref{fig:final_info}; (a) shows the information (equivalently, the logarithm of the posterior) where variations in capsule dimensions and atomic composition are neglected, and (b) show the results when they are included. Both plots use an uncorrelated normal distribution for the prior, with a standard deviation of 0.1 in both directions; as such both represent a more advanced analysis than the $\chi^{2}$ approach shown in figure \ref{fig:normal_likelihood_info}. The distortion of the likelihood by nuisance parameters has the effect that the peak in the information is more localised, however the overall distribution is broader suggesting a larger error bar on the inferred values.\par
\begin{figure}
		\centering
		\includegraphics[scale=1.0]{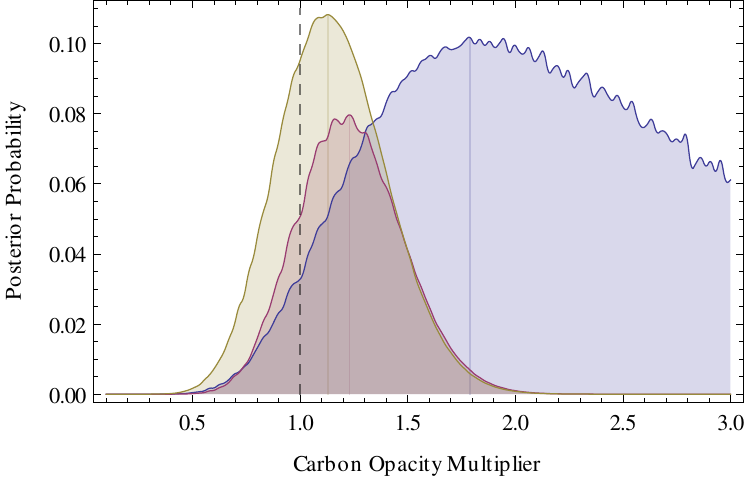}
		\caption{The posterior probability of the carbon opacity multiplier, calculated neglecting nuisance parameters (purple), and including them (yellow). Also shown are the results when a standard $\chi^{2}$ analysis is used (blue). Variations in target properties cause a shift in the maximum likelihood estimate of the opacity multiplier, and a change in the shape of the distribution of values.}
		\label{fig:posterior_c_opac}
	\end{figure}

This can be made more clear by re-casting the data as the posterior probability of the value of the carbon opacity multiplier, by integrating over the drive multiplier. This is not equivalent to the treatment of the drive multiplier as `nuisance'; the detailed response of the system to changes in drive has been included (not the linear response). This approach will prove more accurate for some rapidly varying parameters, however represents a calculation time penalty. The advantage is in the inspection of correlations between parameters that the multi-dimensional analysis allows. The results of integrating over variations in X ray drive are shown in figure \ref{fig:posterior_c_opac}, with the blue line showing the inference when nuisance parameters are included, and purple showing the results when they are neglected. There is a significant change in the posterior probability with a corresponding change in the inferred carbon opacity modifier. Table \ref{tab:inferred_c_opacity} shows the results of this inference for the different Bayesian models we have discussed. These results are (weakly) sensitive to the choice of prior; in cases where the priors are taken as centred on 1, the inferred carbon opacity multiplier is never larger than $1.2$.\par
	\begin{table}
		\centering
		\begin{tabular}{|c|c|}
			\hline
			Inference Model		&	C opacity Multiplier	\\
			\hline
			\hline
			$\chi^{2}$		&	$1.79 \mathsmaller{^{+0.79}_{-0.46}}$		\\
			+ prior			&	$1.23 \mathsmaller{^{+0.23}_{-0.25}}$		\\
			+ nuisance parameters.	&	$1.13 \mathsmaller{^{+0.26}_{-0.25}}$		\\
			\hline
		\end{tabular}
		\caption{The results of the inference of the carbon multiplier from measured values of the implosion velocity, radius and fraction of ablator remaining in NIF convergent ablator shot N110625. Error bars are 68\% confidence limits. Inference is based on a database of 1024 post-shot HYDRA simulations, using the Bayesian inference models described in the text. The inclusion of prior information and of the variation in capsule dimensions and chemical composition is shown to signficantly reduce the measured carbon opacity multiplier, and the error bar on the result.}
		\label{tab:inferred_c_opacity}
	\end{table}
\section{Analysis of a High-Dimensional Parameter Space}
The complexity of ICF experiments, and of the hydrodynamic simulations used to analyse and design them, make it unfeasible to perform the simple grid parameter scans that have concerned us until now. The statistical framework does, however, allow more advance Monte-Carlo approaches to be applied, and these have seen great success in the analysis of other large experiments. The application to ICF experiments will still require careful consideration, however, and in the following we will discuss the possible approaches that are available.\par 
\subsection{Linear Response} \label{sec:linear_response}
The linear response model described above allows the inclusion of nuisance parameters as described above, however the dependance on parameters that are of experimental interest should be treated more accurately. There is an advantage to keeping as many simulation parameters as possible out of the linear response model; as we have seen in the previous low dimensional analysis, inspection of the posterior as a function of the model parameters can provide valuable insight. It is not possible to include all parameters however, and so the linear response model will be essential in the implementation of the following advanced methods.\par
\subsection{Markov Chain Monte Carlo}
The MCMC approach is a statstical method of generating a set of samples of parameter space such that the distribution of samples reflects a given, unknown, probability distribution \cite{green95}. The method has been used extensively to probe posterior distributions with very large numbers of parameters; the potential pitfall is that using HYDRA simulations to form the step probability in these models, even using the efficient routines applied here, presents a significant computational challenge.\par
\begin{figure}
		\centering
		\includegraphics[scale=1.0]{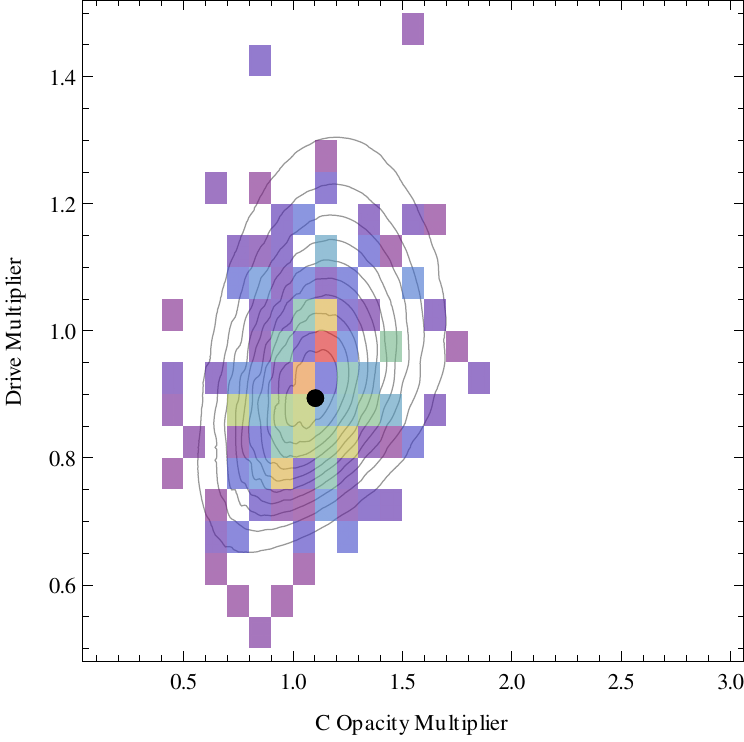}
		\caption{Results of an MCMC simulation of the posterior distribution of carbon opacity and drive multipliers for NIC shot N110625. Contours show the analytically derived posterior, and colored blocks show a histogram of the MCMC results. Nuisance parameters are included using a linear response likelihood, and HYDRA results are interpolated from the simulation grid described in the text.}
		\label{fig:mcmc_posterior}
	\end{figure}

The feasability of the MCMC approach can be investigated using a surrogate model, that will approximately reproduce the multi-dimensional behaviour of HYDRA simulations. In this way the number of steps required can be found, before resorting to time-consuming HYDRA simulations. An example solution is shown in figure \ref{fig:mcmc_posterior}, where we plot a histogram of MCMC samples from a run of 2000 steps (initialised at the peak of the prior and with a 500 step `burn in' period). These simulations use the Metropolis-Hastings algorithm with a normally distributed jump probability to explore the posterior probability formed from the linear response likelihood and prior described in previous sections. The likelihood is approximated by interpolating in the simulation grid described previously, with very good results.\par
It is well known \cite{sivia} that even when MCMC simulations poorly describe the multi-dimensional target distribution, low dimensional quantities are often well described. This is seen in our results where the MCMC average and standard deviation carbon opacity multiplier are $1.10\pm0.23$; for an MCMC run with 1000 (post burn in) samples, this value is $1.09\pm0.22$ which still shows good accuracy.\par
The very small number of simulations required to give a good reproduction of the opacity multiplier is encouraging. It should be noted, however, that even though 2500 HYDRA simulations are easily acheivable using the parrallel control routines used here, MCMC simulations do not lend themselves to easy parrallelisation. The genetic algorithm approach described in the next section are more ameanable to the large scale simulations necessary here.\par
\subsection{Genetic Algorithms}
A GA is a computational tool where a set of samples of parameter space are selected and iterated in order to find the extreme value of some merit function, or fitness. New samples are generated from the old using methods inspired by natural selection - those members that have a large fitness are preferentially selected, and bred together. Random jitter is introduced to avoid local maxima in the global fitness, mimiking random mutation in the current population. In data analysis applications it is usual to use the inverse $\chi^{2}$ function as a fitness, and in this application genetic algorithms are a very robust tool for finding the best fit to data when the parameter space has a high dimension. In previous sections we noted that the usual $\chi^{2}$ function can be interpreted as the information in the posterior, and showed how this can be modified to include variations in nuisance parameters and prior information. These considerations allow the application of genetic algorithm techniques to our Bayesian analysis.\par 
 	\begin{figure}
		\centering
		\includegraphics[scale=1.0]{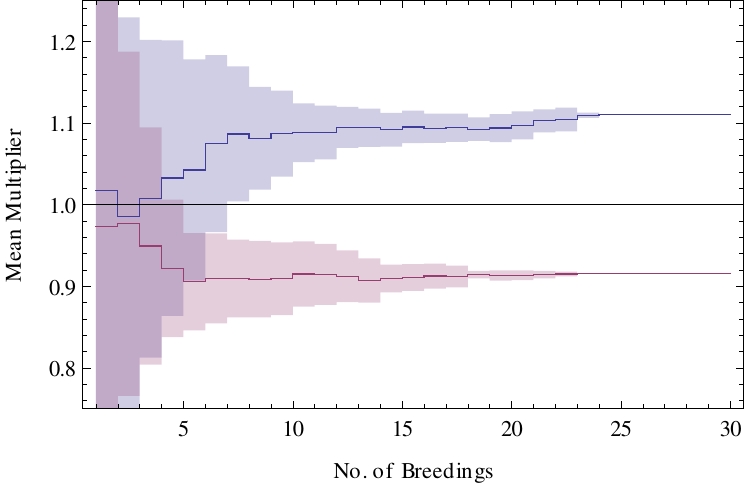}
		\caption{The population mean and standard deviation carbon opacity (blue) and drive (red) multipliers as a function of genetic algorithm breedings. In this simulation a population of 100 simulations were used at each step; this graph therefore represents 2000 HYDRA simulations, which were in this case simulated using an interpolation of a grid of simulations}
		\label{fig:ga_meanC}
	\end{figure}
In figure \ref{fig:ga_meanC} we plot the evolution of the mean carbon opacity and drive multipliers of a population of 100 parameter space samples as subsequent generations are generated. We use a simple GA, where half the population is carried over between generations and the entire population is allowed to mutate. The breeding population is chosen according to the inverse of the generalised $\chi^{2}$ discussed previously. The carbon opacity approaches the expected value of 1.10 quickly, a value that is quite robust as the number of population members (and therefore HYDRA calculations) is reduced. In the case considered here a full GA simulation would require less than 2000 HYDRA simulations, representing a modest computational problem.\par
The genetic algorithm technique is very easily parrallelised since all members of the sample population are known at the start of each iteration. It has the disadvantage, however, that the converged result gives no information regarding the form of the posterior distribution and so it is difficult to provide confidence intervals for the inferred parameters, or the information entropy in the posterior. A simple exploration of of the space close to the maximum, in order to calculate the Hessian matrix \cite{sivia}, would provide an efficient solution to the problem.\par
%
%
\section{Discussion and Conclusions}
We have introduced the Bayesian method and developed the normal linear model to describe nuisance parameters in integrated high energy density experiments. This approach allows a modified Bayesian $\chi^{2}$ function to be defined that efficiently includes prior information and nuisance parameters. This function can be easily applied in any data analysis or experimental design problem, and is particularly well suited to systems with a large number of nuisance parameters. The compact format provided by the modified covariance matrix means that once a sensitivity study has been performed for a given experimental design, the results can be encorporated in all subsequent data analyses in a simple and consistent manner. The expression of the likelihood in these terms also allows powerful Bayesian and Information-Theoretic results to be used.\par
These considerations have been applied to a NIF convergent ablator experiment, where it has been shown that they lead to a significant change in the inferred values of the carbon opacity. It has also been shown that the use of computational methods to speed up analysis is very feasible. This capability, along with the Bayesian aspects presented here, will allow the integrate analysis of entire experimental campaigns at the NIF and to the staistical design of future experiments.\par
%
%
\section{Acknowledgments}
The authors would like to thank Robbie Scott and John Castor for help with HYDRA simulations, and Vijay Sonnad for stimulating comments.
%
%
%
	\refstepcounter{section}									
	\addcontentsline{toc}{section}{\protect\numberline{\thesection}References}				
	\bibliographystyle{unsrt}								
	%

%
%
	\end{document}